\documentclass[useAMS,usenatbib]{mn2e}
\usepackage{graphicx}
\usepackage{txfonts}
\usepackage{color}

\let\tilde=\widetilde
\title[A parametric model for galaxy clusters]{A simple
parametric model for spherical galaxy clusters}
\label{firstpage}
\author[Olamaie et~al.]{Malak Olamaie,$^{1}$\thanks
{Email:mo323@mrao.cam.ac.uk} Michael P. Hobson$^{1}$ and  
Keith J. B. Grainge$^{1,2}$\\
$^{1}$ Astrophysics Group, Cavendish Laboratory, 19 J. J. 
Thomson Avenue, Cambridge, CB3 0HE\\
$^{2}$ Kavli Institute for Cosmology Cambridge, Madingley 
Road,Cambridge, CB3 0HA}

\begin{document}

\date{Accepted 2012 March 21. Received 2012 March 2; in original form 2011 September 13}

\pagerange{\pageref{firstpage}--\pageref{lastpage}}

\pubyear{2011}

\maketitle
\begin{abstract}
We present an analytic parametric model to describe the baryonic and
dark matter distributions in clusters of galaxies with spherical
symmetry. It is assumed that the dark matter density follows a
Navarro, Frenk and White (NFW) profile and that the gas pressure is
described by a generalised NFW (GNFW) profile. By further demanding
hydrostatic equilibrium and that the local gas fraction is small throughout
the cluster, one obtains unique functional forms, dependent on basic 
cluster parameters, for the radial profiles of all the properties of 
interest in the cluster. We show these profiles are consistent both with 
numerical simulations and multi-wavelength observations of clusters. We also 
use our model to analyse six simulated SZ clusters as well as A611 SZ data from the Arcminute 
Microkelvin Imager (AMI). In each case, we derive the radial profile of the enclosed total 
mass and the gas pressure and show that the results are in good agreement with our model prediction.
\end{abstract}
\section{Introduction}
\label{sec:intro}

Analyses of observations of galaxy clusters via their X-ray emission,
gravitational lensing or Sunyaev-Zel'dovich (SZ) effect are often based on
some parameterised cluster model for the distribution of the cluster
dark matter and the thermodynamical properties of its intra-cluster
medium (ICM).  The accuracy and robustness of cluster parameters
derived from studies at different wavelengths depend greatly on how
well the model describes the physical properties of the cluster, and
the assumptions made regarding the dynamical state of the cluster and
its gas content.

Cluster models typically assume spherical symmetry, an ideal gas
equation-of-state, and parameterised functional forms for the radial
distribution of two linearly-independent cluster properties, such as
electron density and temperature (Sanderson \& Ponman 2003; Vikhlinin et~al. 2005 , 2006; Laroque
et~al. 2006; Feroz et~al.  \ 2009; AMI Consortium: Zwart et~al.\ 2011; AMI Consortium:
Rodr\'{i}guez-Gonz\'{a}lvez et~al.\ 2011; AMI Consortium:
Hurley-Walker et~al. 2012 and AMI Consortium: Shimwell et~al. 2011  ); electron pressure and density
(Nagai et~al.\ 2007; Mroczkowski et~al.\ 2009; Arnaud et~al.\ 2010;
Plagge et~al.\ 2010 and Planck Collaboration 2011d); or electron
pressure and entropy (Allison et~al.\ 2011; AMI Consortium: Olamaie
et~al.\ 2012). Such parameterisations are usually supplemented by the
imposition of a condition on the dynamical state of the cluster, most
commonly hydrostatic equilibrium or a virial relation, together
sometimes with further constraints, such as a constant gas fraction
throughout the cluster and/or assorted scaling relations.  All such
models, with their corresponding assumptions, have the potential to
introduce biases in the derived cluster physical parameters, depending
strongly on the appropriateness of the assumptions made and whether or
not the data can constrain the parameters describing the model (see AMI 
Consortium: Olamaie et~al. 2012).

A recent interesting example of a parameterised spherical cluster
model by Mroczkowski  (2011) assumes the cluster dark matter
density to follow a parameterised Navarro, Frenk and White (NFW)
(Navarro et~al. 1997) profile and the gas pressure to be described by
a generalised NFW (GNFW) profile (Nagai et~al. 2007) with fixed shape
parameters, both in accordance with numerical simulations.  This model
also assumes hydrostatic equilibrium and, crucially, a constant gas
fraction (both local and enclosed) throughout the cluster, which is a very stringent condition.
In this paper, we adapt this model by replacing this last condition,
which is in fact inconsistent with the rest of the model, by the much
weaker assumption that the local gas fraction throughout the cluster 
is small compared with unity. We show that this assumption leads to a
unique solution for the radial dependence of all the cluster
properties of interest (dependent on basic cluster parameters). 

Further, we analyse six 
simulated clusters and one real cluster (A611) in this frame work through their Sunyaev--
Zel'dovich effect and show that the resulting profiles agree with those predicted by 
numerical simulations and measured in multi-wavelength observations of galaxy clusters.
\section{The Model}
\label{sec:model}
The first assumption in our model is a functional form for the dark
matter density $\rho_{\rm {DM}}(r)$.  Cosmological $N$-body
simulations suggest that all dark matter halos can be modelled with
the spherically averaged density profile of Navarro, Frenk and White
(NFW) (Navarro et~al. 1997)
\begin{equation}\label{eq:DMdensity}
\rho_{\rm {DM}}(r)=\frac{\rho_{\rm {s}}}{\left(\frac{r}{R_{\rm s}}\right)\left(1 + \frac{r}{R_{\rm s}}
\right)^2},
\end{equation}
where $\rho_{\rm {s}}$ is an overall normalisation coefficient and
$R_{\rm s}$ is the scale radius where the logarithmic slope of the
profile ${\rm d}\ln \rho(r)/{\rm d}\ln r=-2$. It is common practice
also to define the halo concentration parameter,
$c_{200}=\frac{r_{200}}{R_{\rm s}}$, where $r_{200}$ is the radius at
which the enclosed mean density is $200$ times the critical density at the
cluster redshift.

Our second assumption is a functional form for the gas pressure
$P_{\rm gas}(r)$.  Numerical simulations (Nagai et~al. 2007) and X-ray
observations of clusters of galaxies using \textit{Chandra} (Vikhlinin et~al. 2005, 2006; 
Nagai 2006; Nagai et~al. 2007 ) both show that self-similarity is
more likely to be observed in the gas pressure profile than the
density or temperature at large radii, i.e. up to $r_{500}$ and beyond. The gas 
pressure is also the
quantity least affected by dynamical history and non-gravitational
mechanisms inside the ICM. In particular, following Nagai
et~al. (2007), we assume the electron pressure follows the GNFW
profile
\begin{equation}\label{eq:GNFW}
P_{\rm e}(r)=\frac{P_{\rm {ei}}}{\left(\frac{r}{r_{\rm p}}\right)^c\left(1+\left(\frac{r}{r_{\rm
p}}\right)^{a}\right)^{(b-c)/ a}},
\end{equation}
where $P_{\rm {ei}}$ is an overall normalisation coefficient, $r_{\rm
  p}$ is the scale radius. It is common to define the latter in terms
of $r_{\rm 500}$, the radius at which the mean enclosed density is 500 
times the critical density at the cluster redshift, and the gas
concentration parameter, $c_{\rm 500}=r_{\rm 500}/r_{\rm p}$. The
parameters $(a,b,c)$ describe the slopes of the pressure profile at
$r\approx r_{\rm p}$, $r> r_{\rm p}$ and $r \ll r_{\rm p}$
respectively. In the simplest case, we follow Arnaud et~al. (2010) and
fix the values of the gas concentration parameter and the slopes to be
$(c_{\rm 500},a,b,c)=(1.156,1.0620, 5.4807, 0.3292)$. Arnaud et~al. (2010)
derived these values by analysing profiles of the REXCESS
cluster sample observed with XMM-Newton (B\"ohringer et~al. 2007) as well as three different
sets of detailed numerical simulations by Borgani et~al. (2004),
Piffaretti \& Valdarini (2008) and Nagai et~al. (2007) that take into
account radiative cooling, star formation, and energy feedback from
supernova explosions. They estimated $M_{500}$ for each cluster in their sample using the 
standard $M_{500}-Y_x$ scaling relation (see Appendix B in Arnaud et~al. 2010). It should be 
noted that these values are different from those used by Nagai 
et~al. (2007), Mroczkowski et~al. (2009), Plagge et~al. (2010) and Mroczkowski (2011). The 
Arnaud values were, however, used to analyse SZ data from the Planck survey data (Planck 
Collaboration 2011d). 

More generally, one can allow the parameters $(c_{\rm
  500},a,b,c)$ to vary, although we will not consider this case here.
Given the electron pressure, the gas pressure is then defined by
\begin{equation}\label{eq:Pgas}
P_{\rm {gas}}(r)=\frac{\mu_{\rm e}}{\mu}P_{\rm {e}}(r),
\end{equation}
where $\mu_{\rm e}=1.14m_{\rm p}$ (Mason \& Myers
2000) is the mean gas mass per electron, $\mu=0.6m_{\rm p}$ is the
mean mass per gas particle and $m_{\rm p}$ is the proton mass. 

Our third assumption concerns the dynamical state of the cluster,
which we take to be in hydrostatic equilibrium throughout. Thus, the
total cluster mass internal to radius $r$ is related to the gas
pressure gradient at that radius by
\begin{equation}\label{eq:HSE}
\frac{{\rm d}P_{\rm{gas}}(r)}{{\rm d}r} = -\rho_{\rm
  {gas}}(r)\frac{{\rm G}M_{\rm {tot}}(r)}{r^2}.
\end{equation}
However, we note that the latest cosmological simulations of galaxy 
clusters with focus on studying the cluster outskirts (Lau et~al. 2009; Battaglia et~al. 
2010; Nagai 2011; Nagai \& Lau 2011; Parrish et~al. 2012 and Battaglia et~al. 2011 a,b) and 
observational studies of the clusters using the \textit{Suzaku} and XMM- Newton 
satellites at large radii- out to the virial radius including A1795 (Bautz et~al. 2009), PKS 
0745-191 (George et~al. 2009), A2204 (Reiprich et~al. 2009), A1413 (Hoshino et~al. 2010), 
A1689 (Kawaharada et~al. 2010), Virgo cluster (Urban et~al. 2011) and Perseus cluster 
(Simionescu et~al. 2011), show that the presence of random gas motion, gas clumping and 
turbulence due to the magnothermal instability in the intracluster medium of galaxy clusters 
provide non-thermal pressure support and can introduce biases in HSE measurements of the ICM 
profiles and cluster mass. Hence in order to recover these profiles accurately we need to 
modify the equation of HSE to take into account non-thermal pressure components. However, as the 
studies of this kind (to understand the physics of the cluster outskirts and make accurate 
measurements of the ICM profiles in the cluster outer regions) are still ongoing, we do not 
study a modified form of HSE here. We , of course, aim to consider a more general form in our 
future analyses.

Finally, our model is completed by assuming that the local gas
fraction is much less than unity throughout the cluster, i.e. $ \frac{\rho_
{\rm gas}(r)}{\rho_{\rm tot}(r)}\ll 1$ for all $r$. This final assumption allows us to
write $\rho_{\rm tot}(r) = \rho_{\rm DM}(r) + \rho_{\rm gas}(r)
\approx \rho_{\rm DM}(r)$. We emphasize that this assumption is for 
the gravitational part of the calculation and as we show in equation (\ref{eq:rhogas}) we do 
not assume $\rho_{\rm gas}(r)=0$ . Thus, from (\ref{eq:DMdensity}), the total
mass enclosed with a radius $r$ has the analytical solution
\begin{eqnarray} \label{eq:DMmass}
M_{\rm {tot}}(r)&=& \int_{0}^{r}{\rho_{\rm {DM}}(r')(4\pi r^{'2}{\rm d}r')}   \nonumber\\
&=& 4\pi \rho_{\rm {s}}R^3_{\rm s} \left \{\ln\left(1 +\frac{r}{R_{\rm s}}\right)- \left(1+
\frac{R_{\rm s}}{r}\right)^{-1}\right\}.
\end{eqnarray}
Substituting this form and the expressions (\ref{eq:GNFW}) and
(\ref{eq:Pgas}) for the gas pressure into the condition (\ref{eq:HSE})
for hydrostatic equilibrium, one may derive the gas density profile
\begin{eqnarray}\label{eq:rhogas}
\rho_{\rm {gas}}(r) & = & \left(\frac{\mu_{\rm e}}{\mu}\right)\left(\frac{1}{4\pi {\rm G}}
\right)\left(\frac{P_{\rm{ei}}}{\rho_
{\rm {s}}}\right)\left(\frac{1}{R^3_{\rm s}}\right)\times     \nonumber\\
 &  &  \frac{r}{\ln\left(1 +\frac{r}{R_{\rm s}}\right)- \left(1+\frac{R_{\rm s}}{r}\right)^
{-1}}\times \nonumber\\
&  &  \left(\frac{r}{r_{\rm p}}\right)^{ {-c}}\left[1 + \left(\frac{r}{r_{\rm p}}\right)^
{ a}\right]^{-\left(\frac{{ {a + 
b - c}}}{{ a}}\right)}\left[{ b} \left(\frac{r}{r_{\rm p}}\right)^{ a} + {c} \right]
\end{eqnarray}
The radial profile of the electron number density is then trivially
obtained using $n_{\rm e}(r)= \rho_{\rm {gas}}(r)/\mu_{\rm
  e}$. Assuming an ideal gas equation of state, this in turn yields
the electron temperature profile ${\rm k_{\rm B}} T_{\rm{e}}(r) =
P_{\rm e}(r)/n_{\rm e}(r)$, given by
\begin{eqnarray}\label{eq:Tgas}
{\rm k_{\rm B}}T_{\rm{e}}(r) & = & (4\pi \mu {\rm G}\rho_{\rm {s}})(R^3_{\rm s})\times \nonumber\\
 &  &  \left [ \frac{\ln\left(1 +\frac{r}{R_{\rm s}}\right)- \left(1+\frac{R_{\rm s}}{r}\right)^{-1}}{r}  \right] \times
\nonumber\\
 &  &  \left [1 + \left(\frac{r}{r_{\rm p}}\right)^{ a} \right]\left[{b} \left(\frac{r}{r_{\rm p}}\right)^{a} + {c} 
\right]^{-1}
\end{eqnarray}
which is also equal to the gas temperature profile ${\rm k_{\rm B}}
T_{\rm{gas}}(r)$. We can also determine the radial profile for
electron entropy of the ICM. In the astronomy literature, for an
adiabatic monatomic gas, entropy is defined as $K_{\rm e}= {\rm k_{\rm
    B}}T_{\rm{e}}(r)n^{-2/3}_{\rm e}(r)$ which is related to the true
thermodynamic entropy per gas particle via $S=\frac{3}{2}k_{\rm
  B}\ln(K_{\rm e})+S_0$ where $S_0$ is a constant (Voit 2005).

The only fundamental cluster property for which the radial profile
cannot be expressed in an explicit analytical form is the gas mass
enclosed within radius $r$, 
\begin{equation}\label{eq:Mgas}
M_{\rm {gas}}(r)= \int_{0}^{r}{\rho_{\rm {gas}}(r^\prime)(4\pi r^{\prime 2}{\rm d}r^\prime)}. 
\end{equation}
For the gas density profile in (\ref{eq:rhogas}), we have been unable
to evaluate this expression analytically, and so $M_{\rm gas}(r)$
must be obtained using numerical integration. Consequently, the enclosed gas
mass fraction profile $f_{\rm {gas}}(r)=M_{\rm {gas}}(r)/M_{\rm {tot}}(r)$ also
cannot be written in closed form. It is  clear, however, that the resulting $f_{\rm {gas}}
(r)$ will not be constant. Therefore, this contradicts the assumption of Mroczkowski (2011) 
of $f_{\rm {gas}}(r)$ being constant. In the next section, we represent the profile of $f_
{\rm {gas}}(r)$ which illustrates this point.
\section{Illustration of cluster properties}
\label{sec:analysis}

In the simplest case, where $a$, $b$ , $c$ and $c_{500}$ in (\ref
{eq:GNFW}) have
fixed values, our cluster model depends only on three parameters:
$\rho_{\rm s}$ and $R_{\rm s}$ in the NFW dark matter density profile
(\ref{eq:DMdensity}) and ${P_{\rm {ei}}}$ in the
pressure profile (\ref{eq:GNFW}). One is, however, free to choose
alternative parameters to define a cluster, although this choice
and the priors imposed on the parameters can lead to very different
results in the analysis of cluster observations (AMI Consortium:
Olamaie et al.\ 2012). Here we will define clusters in terms of the
parameter set $M_{\rm tot}(r_{200})$, $c_{200}$, $f_{\rm
  gas}(r_{200})$ and the redshift $z$, and investigate the resulting
radial profiles of quantities of interest in our cluster model.

For illustration purposes, we will consider clusters at a fixed
redshift $z=0.3$. We will further assume that $f_{\rm
  gas}(r_{200})=0.12$, which is reasonable since we expect the gas mass
fraction to approach the universal baryon fraction at large scales
near the virial radius (Komatsu et~al. 2011; Larson et~al. 2011). We
will consider a selection of 15 clusters equally spaced in the mass
range $1.0\times 10^{14}\, {\rm M}_\odot<M_{\rm
  {tot}}(r_{200})<1.5\times 10^{15}\, {\rm M}_\odot$.  For each
cluster, we also consider $15$ values of $c_ {200}$ in the range
$4{-}6$, since the concentration parameter shows a clear dependence on
the halo mass, with massive halos having lower concentration parameter
(Pointecouteau et~al. 2005; Salvador-Sol\'{e} et~al. 2007; Vikhlinin et~al. 2006). 
Throughout, we also assume a $\rm{\Lambda CDM}$ cosmology with $\, \Omega_{\rm M}=0.3 \, , \, \Omega_{\rm \Lambda}=0.7\, , \, \sigma_{\rm 8}=0.8\, ,\, h =0.7\, ,\, w_{\rm 0}=-1\, ,\, w_{\rm a}=0$.

To determine the radial profiles of quantities of interest for a given
cluster, one must first determine the values of the model parameters $\rho_{\rm s}$, $R_{\rm s}$, $r_{\rm p}$ and ${P_{\rm {ei}}}$. Since $M_{\rm {tot}}(r_{200})$ is
the total amount of matter internal to radius $r_{200}$, one may write
\begin{equation}\label{eq:sphmass}
M_{\rm {tot}}(r_{200})= \frac{4\pi}{3}r^3_{200}(200 \rho_{\rm {crit}}(z)).
\end{equation}
Thus, for a given $M_{\rm {tot}}(r_{200})$ and $z$, one may calculate
$r_{200}$, and hence $R_{\rm s} = r_{200}/c_{200}$.  The value of
$\rho_{\rm s}$ is then obtained by equating the input value of $M_{\rm
  tot}(r_{200})$ with the RHS of (\ref{eq:DMmass}) evaluated at
$r=r_{200}$, and is given by
\begin{equation}\label{eq:rhos}
\rho_{\rm {s}}=\frac{200}{3}\left(\frac{r_{200}}{R_{\rm s}}\right)^3\frac{\rho_{\rm {crit}}(z)}{\left \{\ln\left(1 +\frac{r_
{200}}{R_{\rm s}}\right)- \left(1+\frac
{R_{\rm s}}{r_{200}}\right)^{-1}\right\}}.
\end{equation}
By equating equations (\ref{eq:DMmass}) and  (\ref{eq:sphmass}) at $r_{500}$, one may calculate $r_{500}$ and hence $r_{\rm p}=r_{500}/c_{500}$. Finally, $P_{\rm {ei}}$ is obtained by substituting (\ref{eq:rhogas})
into (\ref{eq:Mgas}), evaluating the RHS at $r=r_{200}$ and equating
the result to $f_{\rm gas}(r_{\rm 200})M_{\rm tot}(r_{200})$. This yields
\begin{eqnarray}\label{eq:Pei}
  P_{\rm {ei}} &=& \left(\frac{\mu}{\mu_{\rm {e}}}\right)({\rm G}\rho_{\rm {s}}R^3_{\rm s}) M_{\rm {gas}}(r_{200})\times 
\nonumber\\
  &  & \frac{1}{
   {\displaystyle \int_{0}^{r_{\rm 200}}} r^{'3} {\rm d}r'
   \frac{\left[b \left(\frac{r'}{r_{\rm p}}\right)^a + c \right]}{ 
   \left[\ln\left(1 +\frac{r'}{R_{\rm s}}\right)- \left(1+\frac{R_{\rm s}}{r'}\right)^{-1}\right]   
   \left(\frac{r'}{r_{\rm p}}\right)^c 
   \left[1 + \left(\frac{r'}{r_{\rm p}}\right)^a\right]^{\left(\frac{{a + b - c}}{a}\right)}
    }           },
\end{eqnarray}
which must be evaluated numerically.

\begin{figure}
\includegraphics[width=80mm]{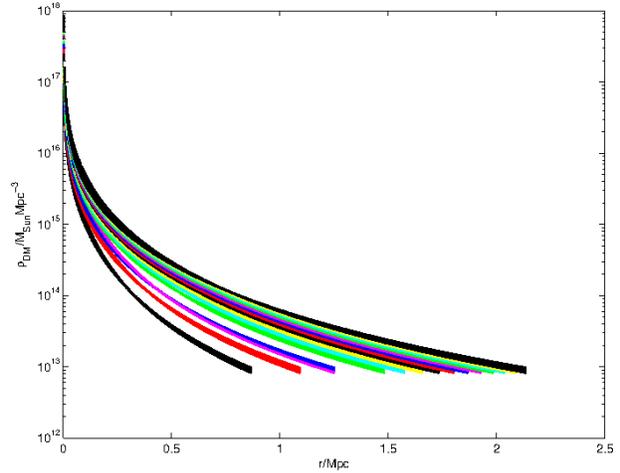}
\caption{Dark matter density profiles $\rho_{\rm DM}(r)$. For a given thick 
line, the thickness represents the spread in varying the halo concentration 
parameter,$c_{200}$ between $4$ and $6$.}  \label{fig:rhoDM}
\end{figure}

\begin{figure}
\includegraphics[width=80mm]{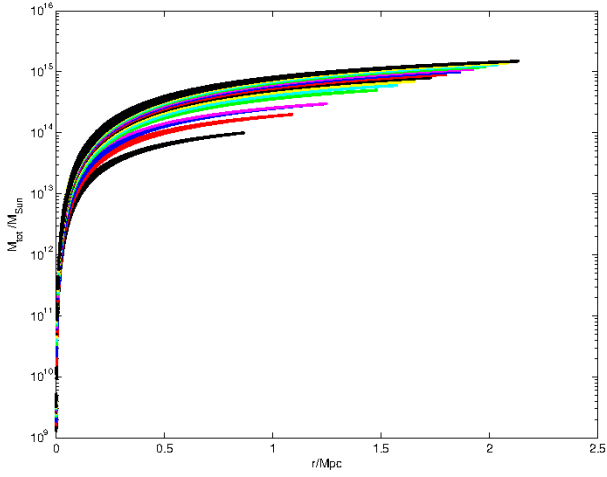}
\caption{Integrated total mass profiles $M_{\rm tot}(r)$.} \label{fig:Mtot}
\end{figure}

\begin{figure}
\includegraphics[width=80mm]{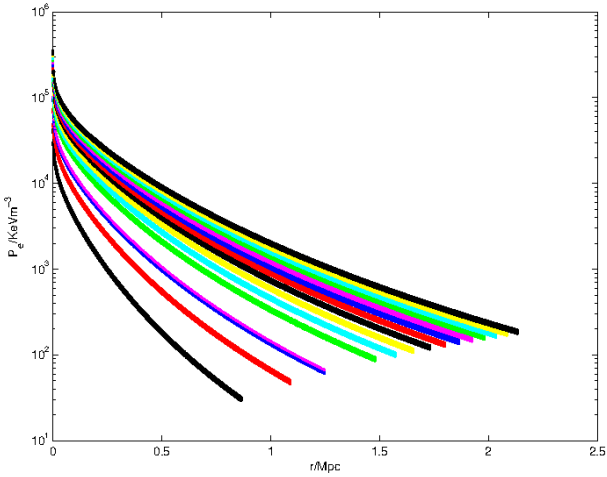}
\caption{Electron pressure profiles $P_{\rm e}(r)$.} 
\label{fig:Pe}
\end{figure}

\begin{figure}
\includegraphics[width=80mm]{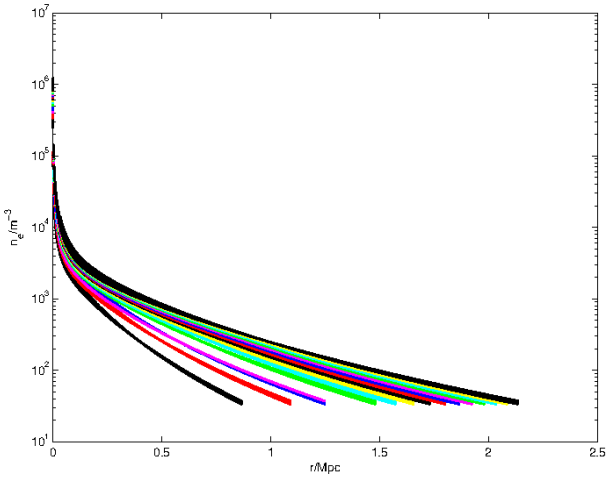}
\caption{Electron number density profiles $n_{\rm e}(r)$.} \label{fig:ne}
\end{figure}

\begin{figure}
\includegraphics[width=80mm]{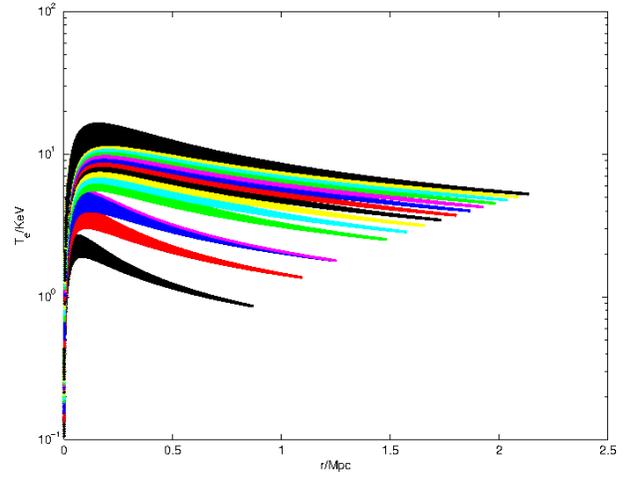}
\caption{Electron temperature profiles $T_{\rm e}(r)$.} \label{fig:Te}
\end{figure}

\begin{figure}
\includegraphics[width=80mm]{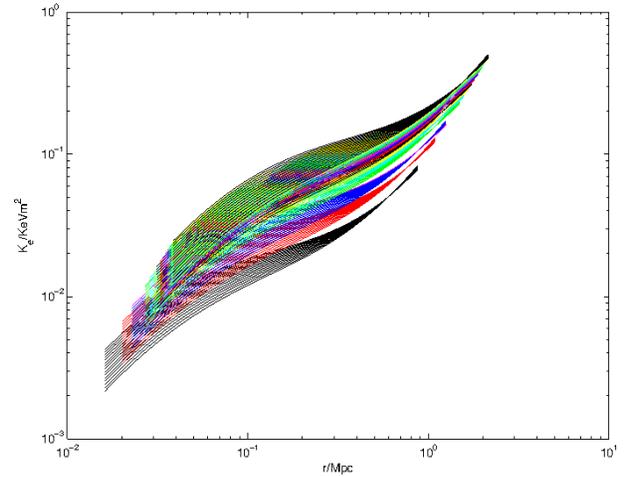}
\caption{Electron entropy profiles $K_{\rm e}(r)$.} \label{fig:Ke}
\end{figure}

\begin{figure}
\includegraphics[width=80mm]{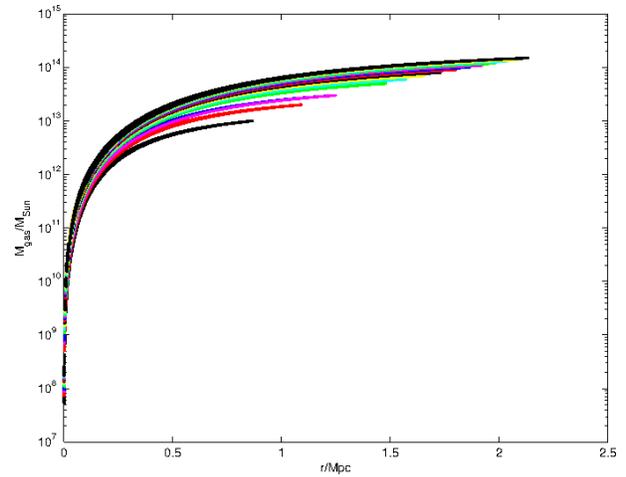}
\caption{Gas mass profiles $M_{\rm gas}(r)$.} \label{fig:Mgas}
\end{figure}

\begin{figure}
\includegraphics[width=80mm]{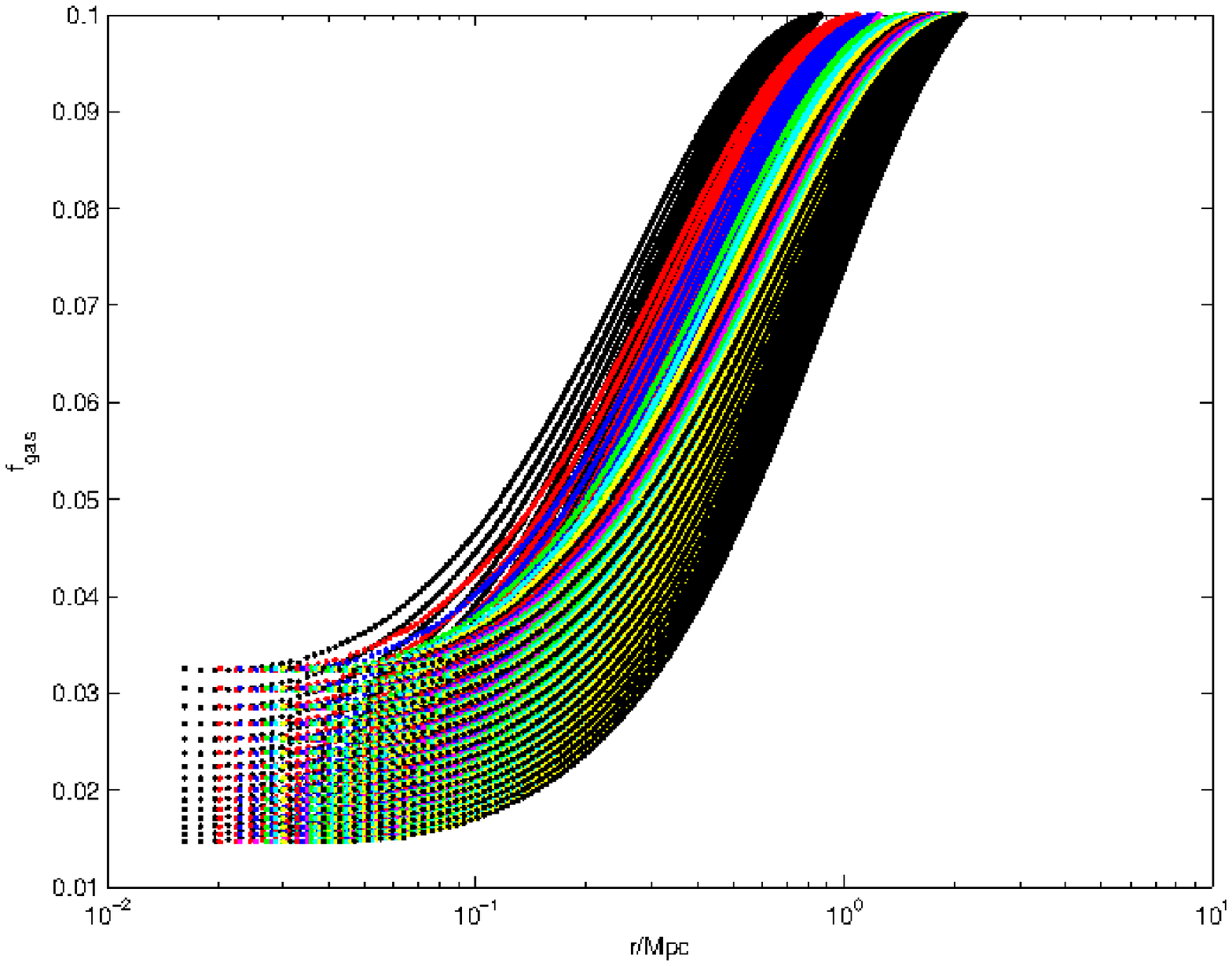}
\caption{Gas mass fraction profiles $f_{\rm gas}(r)$.} \label{fig:fgas}
\end{figure}

Tab.~\ref{tab:pars} summarises the input parameter values for our
illustrative clusters.
\begin{table}
\caption{The input parameters and ranges used in the analysis\label{tab:pars}}
\begin{tabular}{@{}cc@{} }
\hline
Parameter & Value \\\hline
$M_{\rm {tot}}(r_{\rm 200})$  & $(1.0 \,\, \, \, \, 15.0)\times 10^{14}\, \rm{M_\odot}$\\
$c_{200}$ & $(4\,\, \, \,6)$\\
$z$    &   $0.3$               \\
$f_{\rm g}(r_{\rm 200})$  &  $0.12$ \\ \hline
\end{tabular}
\end{table}
The corresponding radial profiles for various quantities of interest
are shown in Figs.~\ref{fig:rhoDM}--\ref{fig:fgas}.  The 
thickness of each line represents the spread in halo concentration parameter corresponding 
to varying $c_{200}$ between $4$ and $6$-i.e.  each
thick line represents clusters with the same $M_{\rm {tot}}(r_{\rm 200})$ but different $c_
{200}$. It should be noted that each profile is plotted out to $r_{\rm 200}$ for the 
corresponding cluster.
%
\section{Analysis of Interferometric SZ Observations}
\label{sec:SZanalysis}
In order to verify that our proposed 
model, with its corresponding assumptions, can describe profiles of 
cluster physical properties accurately, we 
carried out a Bayesian analysis (Feroz \& Hobson 2008; Feroz et~al. 2009a) of 
a set of six simulated clusters as well as A611 observed through their 
Sunyaev-Zel'dovich (SZ) effect (Sunyaev \& Zeldovich 1970; Birkinshaw 1999; 
Calrstrom, Holder \& Reese 2002) using the Arcminute Microkelvin Imager (AMI) 
(AMI Consortium: Zwart et~al. 2008).

The observed SZ surface brightness in the direction of electron reservoir may 
be described as
\begin{equation}\label{deltaI}
\delta I_\nu=T_{\rm CMB}yf(\nu)\frac{\partial B_\nu}{\partial T}\Big\vert_{T=T_
{\rm CMB}}.
\end{equation}
Here $B_\nu$ is the blackbody spectrum, $T_{\rm CMB}=2.73 $~K (Fixsen et~al. 
1996) is the temperature of the CMB radiation, $f(\nu)=\left(x\frac{e^x+1}
{e^x-1}-4\right)(1 + \delta (x , T_{\rm e})$ is the frequency dependence of 
thermal SZ signal, $x=\frac{h_{\rm p}\nu}{k_{\rm B}T_{\rm CMB}}$, $h_{\rm p}$ 
is Planck's constant, $\nu$ is the frequency and $\rm{k_{\rm B}}$ is 
Boltzmann's constant. $\delta (x , T_{\rm e})$ takes into account the 
relativistic corrections in the study of the thermal SZ effect which is 
due to the presence of thermal weakly relativistic electrons in the ICM and is 
derived by solving the Kompaneets equation up to the higher orders 
(Rephaeli~1995, Itoh et~al. 1998, Nozawa et~al. 1998, Pointecouteau et~al. 
1998 and Challinor and Lasenby 1998). It should be noted that at 15 GHz (AMI 
observing frequency) $x= 0.3$ and therefore the relativistic correction, as 
shown by Rephaeli (1995), is negligible for $k_{\rm B}T_{\rm e} \leq 15\, \rm
{keV}$. The dimensionless parameter $y$, known as the Comptonization parameter, is 
the integral of the number of collisions multiplied by the mean fractional 
energy change of photons per collision, along the line of sight
\begin{eqnarray}\label{eq:ypar}
 y &=& \frac{\sigma_{T}}{m_{\rm e}c^2} \int_{-\infty}^{+\infty}{n_{\rm e}(r)k_
{\rm B}T_{\rm e}(r){\rm d}l}\\
&=& \frac{\sigma_{T}}{m_{\rm e}c^2} \int_{-\infty}^{+\infty}{P_{\rm e}(r){\rm 
d}l},
\end{eqnarray}
where $n_{\rm e}(r)$, $P_{\rm e}(r)$ and $T_{\rm e}$ are the electron number 
density, pressure and temperature at radius $r$ respectively. $\sigma_{\rm T}$ 
is Thomson scattering cross-section, $m_{\rm e}$ is the electron mass, $c$ is 
the speed of light and $dl$ is the line element along the line of sight. 
It should be noted that in equation~(\ref {eq:ypar}) we have used the ideal gas 
equation of state.

An interferometer like AMI operating at a frequency $\nu$ measures samples
from the complex visibility plane $\tilde{I}_\nu({\bf u})$. These are given by
a weighted Fourier transform of the surface brightness $I_\nu({\bf x})$, namely
\begin{equation}\label{eq:Ifourier}
\tilde{I}_\nu({\bf u})=\int{A_\nu({\bf x})I_\nu({\bf x})\exp(2\pi i{\bf u\cdot 
x}){\rm d}{\bf x}},
\end{equation}
where ${\bf x}$ is the position relative to the phase centre, $A_\nu({\bf x})$
is the (power) primary beam of the antennas at observing frequency $\nu$
(normalised to unity at its peak) and ${\bf u}$ is the baseline vector in units
of wavelength. 

Further details of our Bayesian methodology, modelling interferometric SZ 
data, primordial CMB anisotropies, and resolved and unresolved 
radio point-source models are given in Hobson \& Maisinger (2002), Feroz \& Hobson
 (2008) and (2009 a,b), AMI Consortium: Davies et~al. (2011) and AMI 
Consortium: Olamaie et~al. (2012).

In generating simulated SZ skies and observing them with a model AMI SA, we 
have used the methods outlined in Hobson \& Maisinger (2002), Grainge et~al.
\ (2002), Feroz et~al. (2009b) and AMI Consortium: Olamaie et~al. (2012). 

Generating a simulated cluster SZ signal using the model described in Sections 
\ref {sec:model} and \ref {sec:analysis} requires the input parameters $M_{\rm 
{tot}}(r_{\rm 200})$, $c_{200}$, $z$ and $f_{\rm {gas}}(r_{200})$ listed in Tab.~\ref
{tab:pars}; this set of parameters fully describes the Comptonization $y$ 
parameter. Tab.~\ref{tab:simclpars} summarises the input parameters used 
to generate six simulated SZ clusters. The simulated clusters all have the same  $M_{\rm 
{tot}}(r_{\rm 200})$, $z$ and $f_{\rm {gas}}(r_{200})$ and the only parameter that varies 
from cluster to cluster is the halo concentration parameter, $c_{200}$.

A611 is a rich cluster at redshift $z=0.288$ and has been studied through its X-ray 
emission, strong lensing, weak lensing and SZ effect (Schmidt \& Allen 
2007; Romano et~al. 2010; Donnarumma et~al. 2011 and AMI Consortium: Shimwell et~al. 
2011).These studies suggest that there is no significant contamination from radio sources 
and there is no evidence for a radio halo associated with A611 (Venturi et~al. 2008). The 
SZ signal (decrement) on the AMI map appears circular, (fig. 2 in AMI Consortium: Shimwell 
et~al. 2011) in agreement with the X-ray surface brightness from the \textit{Chandra} 
archive data (fig. 2 in AMI Consortium: Shimwell et~al. 2011 and fig. 1 in Donnarumma 
et~al. 2011), which also appears to be smooth and whose peak coincides with the position of 
the brightest cluster galaxy and the SZ peak. These results are a strong indication that 
the cluster is relaxed. Moreover, the absence of radio halos in the cluster which are major 
sources of the presence of non-thermal mechanisms in the galaxy clusters ( Brunetti 
et~al. 2009) makes A611 an ideal cluster candidate for our analysis as it satisfies both 
assumptions of spherical symmetry and thermal pressure support in equation of the HSE. 

Details of AMI pointed observation towards the cluster, data reduction 
pipeline and mapping are described in AMI Consortium: Shimwell et~al. (2011) and in here we 
focus on the Bayesian analysis of the clusters using our model. 

\begin{table}
\caption{The input parameters used to generate simulated clusters \label{tab:simclpars}.}
\begin{tabular}{@{}lcccc@{} }
\hline
Cluster&$M_{\rm {tot}}(r_{\rm 200})\, 10^{14}\,\rm{M_\odot}$& $c_{200}$ &$z$& $f_{\rm g}(r_
{\rm 200})$\\\hline
clsim1&$5.0$&$4.0$&$0.3$&$0.12$\\
clsim2&$5.0$&$4.3$&$0.3$&$0.12$\\
clsim3&$5.0$&$4.5$&$0.3$&$0.12$ \\
clsim4&$5.0$&$5.0$&$0.3$&$0.12$ \\
clsim5&$5.0$&$5.5$&$0.3$&$0.12$ \\
clsim6&$5.0$&$6.0$&$0.3$&$0.12$ \\
 \\ \hline
\end{tabular}
\end{table}
The sampling parameters in our Bayesian analysis are $\mbox{\boldmath$\Theta$}_{\rm c}
\equiv (x_{\rm c}, y_{\rm c}, c_{200} , M_{\rm {tot}}(r_{\rm 200}),f_{\rm g}(r_{\rm 200}), 
z)$, where $x_{\rm c}$ and $y_{\rm c}$ are cluster projected position on the sky. We 
further assume that the priors on sampling parameters are separable (Feroz et~al.\ 2009b) 
such that
\begin{equation}\label{eq:prior}
 \pi(\mbox{\boldmath$\Theta$}_{\rm c})=\pi(x_{\rm c})\,\pi(y_{\rm c})\,\pi(c_{200})\,\pi(M_T
(r_{\rm 200}))\,\pi(f_{\rm g}(r_{\rm 200}))\,\pi(z).
\end{equation}
We use Gaussian priors on cluster position parameters, centred on the pointing centre and 
with standard-deviation of 1 arcmin. We adopt uniform priors on $c_{200}$ and a $\delta$ 
function prior on redshift $z$. The prior on $M_{\rm {tot}}(r_{\rm 200})$ is taken to be 
uniform in log$M$ in the range $M_{\rm {min}} = 10^{14}\,\rm{M_ \odot}$ to $M_{\rm 
{max}} = 6\times10^{15}\, \rm{M_\odot}$ and the prior of $f_{\rm {gas}}(r_{\rm 
200})$ is set to be a Gaussian centred at the  $f_{\rm {gas}}=0.12$ with a width of $0.02$. 
A summary of the priors and their ranges are presented in Tab.~ \ref{tab:clpriors}.
\begin{table}
\caption{Summary of the priors on the sampling parameters. Note that $N(\mu ,\sigma)$ 
represents a Gaussian probability distribution with mean $\mu$ and standard deviation of $
\sigma$ and $U(a,b)$ represents a uniform distribution between $a$ and $b$. \label
{tab:clpriors}}
\begin{tabular}{@{}ll@{} }
\hline
Parameter       &\qquad \qquad Prior                                      \\
\hline
$x_{\rm c}$ , $y_{\rm c}\qquad$ &\qquad \qquad $N(0 \,\, , \, \,60)\arcsec$  \\
$c_{200}\qquad$ &\qquad \qquad $U(1 \,\, , \, \, 10)$ \\
$\log M_{\rm {tot}}(r_{\rm 200})\qquad$ &\qquad \qquad $U(14 \,\, , \, \, 15.8)\,\rm{M_
\odot}$  \\
$f_{\rm {gas}}(r_{\rm 200})\qquad$ &\qquad \qquad $N( 0.12 \,\, , \, \, 0.02)$ \\ \hline
\end{tabular}
\end{table}

\begin{figure}
\includegraphics[width=80mm]{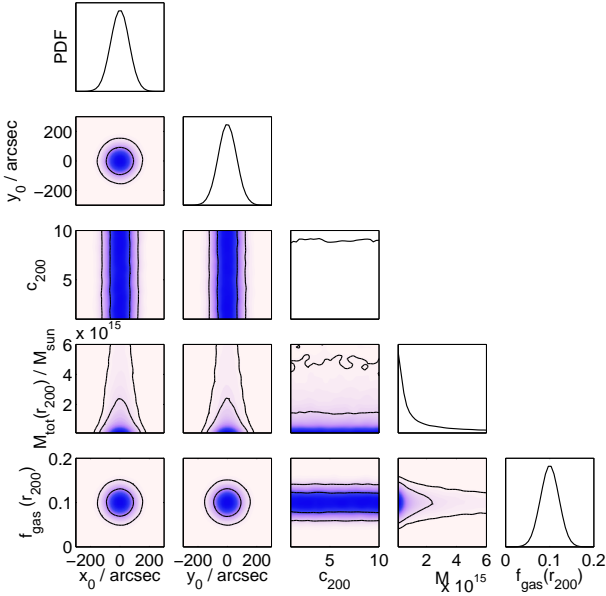}
\caption{1-D and 2-D marginalised posterior distributions of sampling parameters with no 
data \label{fig:nodata}.}
\end{figure}

In order to understand the underlying biases and constraints imposed by the priors and the 
model assumptions, we first study our methodology in the absence of data. This can be 
carried out by setting the likelihood to a constant value and hence the algorithm just 
explores the prior space.  Along with the analysis done using the simulated AMI data, this 
approach reveals the constraints that measurements of the SZ signal place on the cluster 
physical parameters and the robustness of the assumptions made. Fig.~\ref{fig:nodata} 
represents 1-D and 2-D marginalised posterior distributions of a prior-only analysis for each 
of the sampling parameters in our model. The plots show that we correctly recover the 
assumed prior probability distributions of the sampling parameters in the absence of SZ data.

\begin{figure}
\includegraphics[width=80mm]{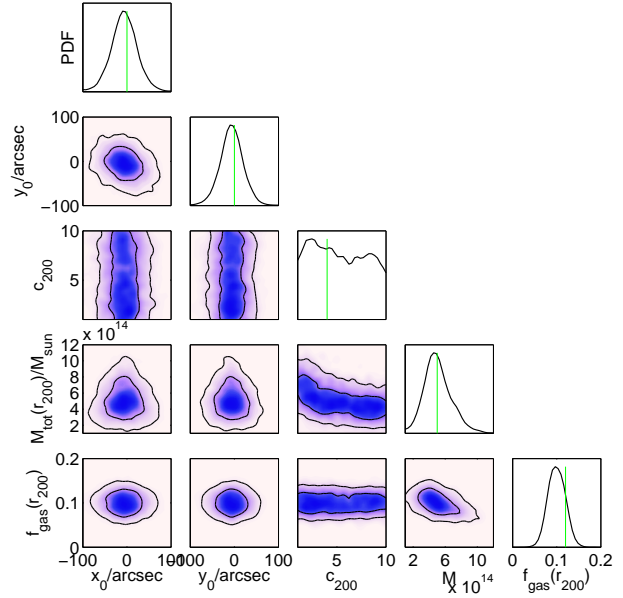}
\caption{1-D and 2-D marginalised posterior distributions of sampling parameters for the 
first simulated cluster \label{fig:simdatatri}.}
\end{figure}

\begin{figure}
\includegraphics[width=80mm]{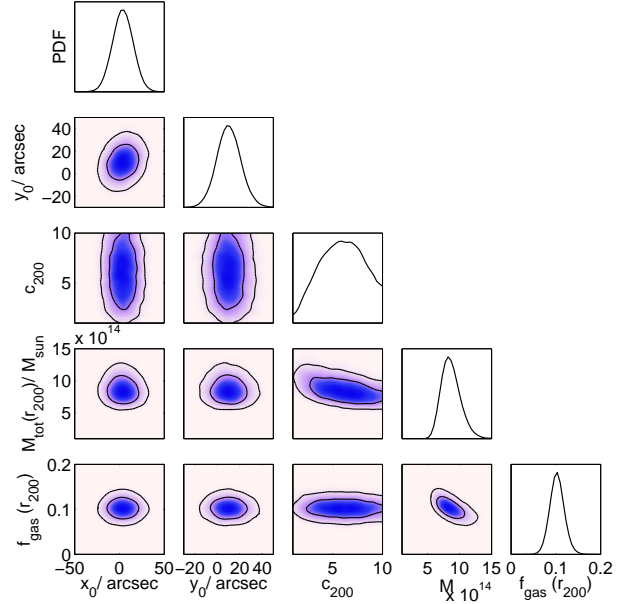}
\caption{1-D and 2-D marginalised posterior distributions of sampling parameters for 
A611 \label{fig:A611datatri}.}
\end{figure}

Fig.~\ref{fig:simdatatri} shows 1-D and 2-D marginalised posterior distributions of 
sampling parameters for the first simulated SZ cluster data, with vertical lines 
representing the true parameter values and Fig.~\ref{fig:A611datatri} shows the results of 
the analysis for A611. From the plots we notice that the model, along with its 
corresponding assumptions, can constrain cluster position and $M_{\rm {tot}}(r_{\rm 200})$, 
but $c_{200}$ remains relatively unconstrained. We also notice the weak negative degeneracy 
between $M_{\rm {tot}}(r_{\rm 200})$ and $c_{200}$ as we expect in high mass halos, between 
$1.0\times10^{14}\,\rm{M_\odot}$ and $15.0\times10^{14}\,\rm{M_\odot}$ (Pointecouteau et~al. 
2005; Salvador-Sol\'e rt~al. 2007; Rudd, Zentner \& Kravtsov 2008; Bhattacharya, Habib, \& 
Heitmann 2011). From our analysis we find $M_{\rm {tot}}(r_{\rm 200})=(5.3\pm 2.6)\times 10^
{14}\,\rm{M_\odot}$ and $r_{200}=(1.5\pm0.2)\, \rm{Mpc}$ for the simulated cluster and  $M_
{\rm {tot}}(r_{\rm 200})=(8.6\pm 1.4)\times 10^{14}\,\rm{M_\odot}$ and $r_{200}=(1.7\pm0.1)\, 
\rm{Mpc}$ for A611.

\begin{figure}
\includegraphics[width=80mm]{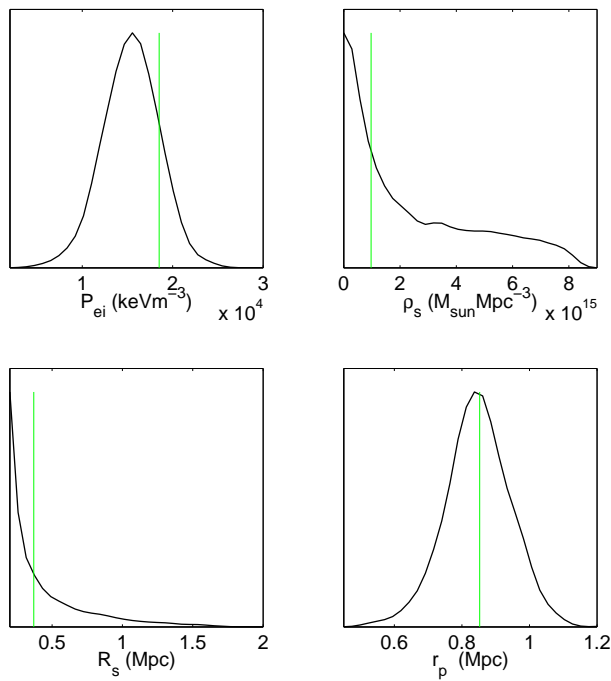}
\caption{1-D marginalised posterior distributions of the model parameters for the first 
simulated cluster. Green vertical lines are the derived values of the model parameters 
for a cluster defined by the input parameters given in the first row of Tab. \ref{tab:simclpars}
\label{fig:1dposnormparssim1}}
\end{figure}

\begin{figure}
\includegraphics[width=80mm]{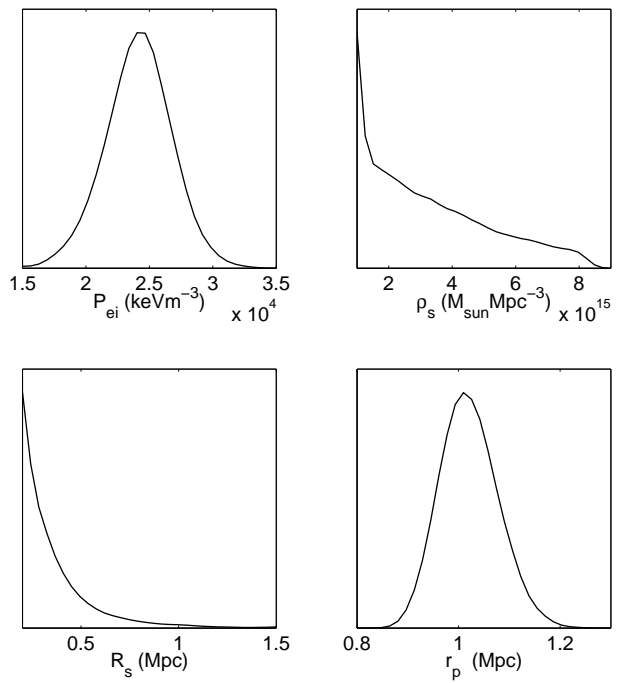}
\caption{1-D marginalised posterior distributions of the model parameters for A611.
\label{fig:1dposnormparsA611}}
\end{figure}

Figs.~\ref{fig:1dposnormparssim1} and \ref{fig:1dposnormparsA611} present 1-D marginalised 
posterior distributions of the model parameters (i.e. $\rho_{\rm s}$, $R_{\rm s}$, $r_{\rm 
p}$ and $P_{\rm {ei}}$) for the first simulated cluster and A611 respectively.  We note that 
although our data can constrain $P_{\rm {ei}}$ and $r_{\rm p}$, $\rho_{\rm s}$ and 
$R_{\rm s}$ are not well constrained as they depend strongly on the relatively unconstrained 
cluster concentration parameter $c_{200}$. We use the best-fit values of these four parameters 
given in Tabs.~\ref{tab:bestfitsimpars} and \ref{tab:bestfitrealpars} to determine the radial 
profiles of the clusters physical parameters according to our model. 
\begin{table}
\caption{Simulated cluster model parameters estimated 
(mean and standard deviation)\label{tab:bestfitsimpars} }
\begin{tabular}{@{}lcc@{} }
\hline
Parameter & $\mu $&$\sigma$  \\\hline
$\rho_{\rm s}\,(\rm{M_\odot\,Mpc^{-3}})$ & $2.52\,\times\, 10^{15}$ & $2.31\,\times\, 10^{15}$ 
\\
$R_{\rm s}\,(\rm{kpc})$ & $402.79$ & $306.83$  \\
$r_{\rm p}\,(\,\rm{kpc})$ & $846.93 $  & $95.88$  \\
$P_{\rm {ei}}\,(\,\rm{keVm^{-3}})$& $1.54\,\times\,10^{4}$ &$3.08\,\times\,10^{3}$             \\
\hline
\end{tabular}
\end{table}
\begin{table}
\caption{Best-fit values of model parameters estimated (mean and standard deviation) 
for A611. \label{tab:bestfitrealpars}}
\begin{tabular}{@{}lcc@{} }
\hline
Parameter & $\mu $&$\sigma$  \\\hline
$\rho_{\rm s}\,(\rm{M_\odot\,Mpc^{-3}})$ & $2.87\,\times\,10^{15}$ & $2.07\,\times\,10^{15}$\\
$R_{\rm s}\,(\rm{kpc})$ & $361.80$ & $209.31$  \\
$r_{\rm p}\,(\rm{kpc})$ & $ 1021.27$  & $55.57$  \\
$P_{\rm {ei}}\,(\,\rm{keVm^{-3}})$& $2.41\,\times\,10^{4}$ &$2.59\,\times\,10^{3}$\\
\hline
\end{tabular}
\end{table}

\begin{figure}
\includegraphics[width=80mm]{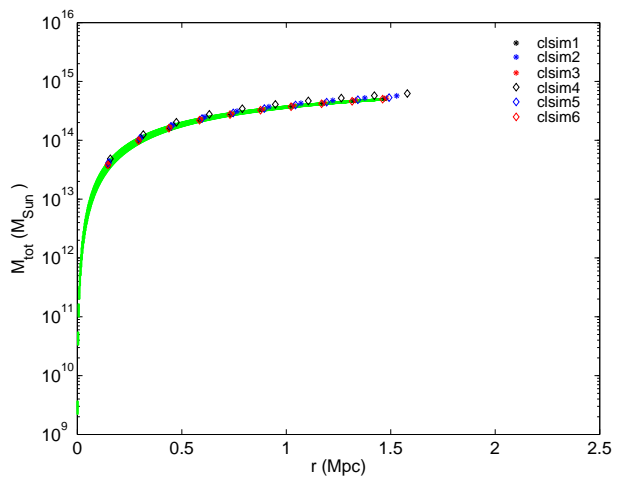}
\caption{Profile of $M_{\rm {tot}}$ versus $r$ for simulated clusters. \label{fig:Mtotsims}}
\end{figure}

\begin{figure}
\includegraphics[width=80mm]{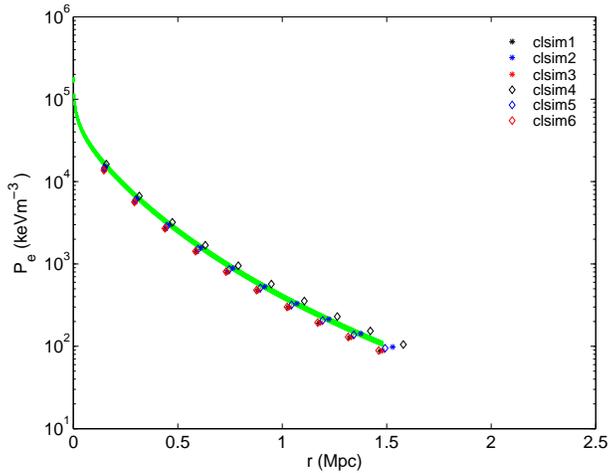}
\caption{Profile of $P_{\rm e}$ versus $r$ for simulated clusters. \label{fig:Pesims}}
\end{figure}

\begin{figure}
\includegraphics[width=80mm]{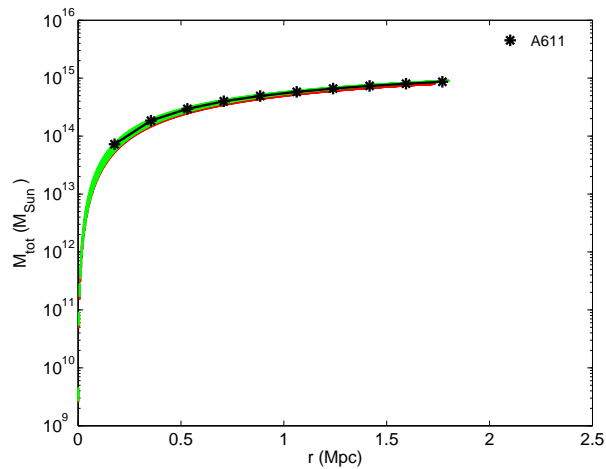}
\caption{Profile of $M_{\rm T}$ versus $r$ for A611. \label{fig:MtotA611}}
\end{figure}

\begin{figure}
\includegraphics[width=80mm]{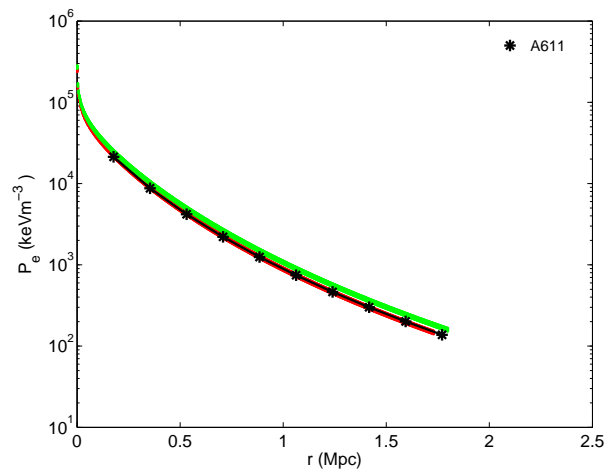}
\caption{Profile of $P_{\rm e}$ versus $r$ for A611. \label{fig:PeA611}}
\end{figure}

As the SZ surface brightness is proportional to the line-of-sight integral of the electron 
pressure, (equations \ref{deltaI} and \ref{eq:ypar}) SZ analysis of galaxy clusters provides 
a direct measurement of the pressure distribution of the ICM. Moreover, the integral of the 
Comptonization $y$ parameter over the solid angle $\Omega$ subtended by
the cluster ($Y_{SZ}$) is proportional to the volume
integral of the gas pressure. It is thus a good estimate for the total thermal
energy content of the cluster and its mass (see e.g. Bartlett \& Silk 1994). $Y_{\rm SZ}$ 
parameter in both cylindrical and spherical geometries may be described as

\begin{eqnarray}\label{eq:Ycylsph}
Y_{\rm cyl}(R)&=& \frac{\sigma_{T}}{m_{\rm e}c^2}\int_{-\infty}^{+\infty}{\rm{d}l}\,
\int_{0}^{R}{P_{\rm e}(r)2\pi s \, \rm {d}s} \\
Y_{\rm sph}(r)&=& \frac{\sigma_{\rm T}}{m_{\rm e}c^2}\int_{0}^{r}{P_{\rm e}(r')4\pi 
r^{'2}\rm {d}r'}  
\end{eqnarray}
where $R$ is the projected radius of the cluster on the sky. In this context we determined the 
radial profiles of $M_{\rm {tot}}$ and $P_{\rm e}$ as a function of $r$ for six simulated 
clusters and A611 (figs. \ref{fig:Mtotsims}-\ref{fig:PeA611}). In all figures, the 
background thick line shows the model prediction of the profiles of the clusters with the same mass as the 
clusters been analysed but vary in $c_{200}$ as was illustrated in figs. ~\ref{fig:rhoDM}--\ref{fig:fgas}. We have plotted the radial profiles 
of the corresponding cluster properties with coloured $\ast$ and $\diamond$ in case of 
simulated clusters and black $\ast$ for A611. From the plots it is obvious that the radial 
trend of the clusters profiles are all consistent with our model prediction. 

%
\section{Discussion and Conclusions}
We have studied the combination of NFW dark matter and GNFW gas
pressure profiles within the hierarchical structure formation scenario
(Kaiser 1986) to derive the radial distribution of the
cluster properties, assuming spherical symmetry, hydrostatic
equilibrium, and that the local gas fraction throughout is small
compared to unity.

Figs.~\ref{fig:rhoDM} and \ref{fig:Mtot} represent the dark matter
density and enclosed mass profiles (the latter approximating the total
enclosed mass profile under our assumption that $ \frac{\rho_
{\rm gas}(r)}{\rho_{\rm tot}(r)}\ll 1$ throughout). These results are based on the assumption 
of an NFW dark matter density profile, arising from the notion that the largest
virialised structures form via gravitational collapse and merging. The
profile has proved to be a good fit to the relaxed dark matter halos
in high resolution N-body simulations down to $1\%$ of the virial
radius, and optical and X-ray observations of galaxy clusters both
indicate that the profile is a good representation of the underlying
cluster mass profile outside the core (Carlberg et~al. 1997; Pratt \& Arnaud 2002).

The electron pressure profiles are shown in Fig.~\ref{fig:Pe}, which
are assumed to follow a GNFW profile. They exhibit self-similarity at
the larger radii as they approach $r_{200}$ and show dependency on the cluster mass. These
behaviours are expected and resemble the profiles observed in a wide
range of clusters (Holder et~al. 2007; Nagai et~al. 2007; Arnaud
et~al.  2010; Plagge et~al.2010; Mroczkowski et~al. 2009) indicating
that the pressure is least affected by non-gravitational phenomena in
the ICM. This is important, in particular, in the analysis of
Sunyaev--Zel'dovich observations of clusters, which essentially
measure the line-of-sight integral of the ICM pressure through the
cluster.

The derived gas density profiles $\rho_{\rm {gas}}(r)=\mu_{\rm
  e}n_{\rm e}(r)$ are shown in Fig.~\ref{fig:ne}, and reproduce all
the main features observed both in numerical simulations and in real
clusters (Sanderson et~al. 2003; Borgani et~al. 2004; Vikhlinin et~al. 2005, 2006; Nagai et~al. 2007; McCarthy et~al. 2008). In particular,
the profiles exhibit steepening at large radii, a power-law cusp at
small radii (resulting from the fact that gas cooling and star
formation processes have been taken into account in deriving the GNFW
pressure profile) and a change of slope at intermediate radii. We
also note that lower temperature clusters have lower gas density. The
derived analytical expression for the gas density can thus model both
the inner and outer regions of the clusters.

The derived electron/gas temperature profiles of the clusters are shown 
in Fig.~\ref{fig:Te}. All of them have similar positive slopes up to $r
\sim 0.1\, \rm{Mpc}$ for the most massive clusters and have a broad
peak around this region.  The significant drop in the temperature in
the innermost region is again because of taking into account the
presence of radiative cooling mechanisms in deriving the GNFW pressure
profile (Borgani et~al. 2004; Vikhlinin et~al. 2005, 2006; Pratt
et~al. 2007; Leccardi \& Molendi 2008). In particular, we note that the
clusters do not have isothermal cores. It should be pointed out,
however, that real cluster data and current high-resolution
simulations display complex temperature structures, which are the
result of merging subgroups or supersonic accretion which heats the
gas across the shock front where the assumption of hydrostatic
equilibrium clearly breaks. Nonetheless, our derived temperature
profile describes the general features of the ICM well, within our
assumptions.

Fig.~\ref{fig:Ke} shows the entropy profiles, which clearly 
show that the entropy depends on temperature and therefore the
cluster mass. Moreover, the entropy profiles approach self-similarity
as the radius approaches $r_{200}$, showing a scaling power-law
distribution which is predicted in the models based on spherical gas
accretion within a NFW dark matter halo (Tozzi \& Norman 2001). This 
demonstrates that gravity dominates the ICM thermodynamics in the outer regions of
clusters. These behaviours have already been noted in the cluster
numerical simulations (Kay et~al. 2004; Borgani et~al. 2004; Voit, Kay
\& Bryan 2005) and have also been observed in a large
sample of galaxy clusters (Ponman, Sanderson and Finoguenov 2003; Piffaretti et~al. 2005; McCarthy et~al. 
2008; Pratt et~al. 2010). In the inner regions, on the other hand, the entropy profiles
are clearly affected by the non-gravitational processes that have been
considered in deriving the GNFW pressure profile.
 
Fig.~\ref{fig:Mgas} presents the enclosed gas mass profiles which  
increasing with radius but with different slopes and fig.~(\ref{fig:fgas}) shows the derived 
gas mass fraction profiles, which also exhibit a significant increase with radius, hence 
implying that $f_{\rm {gas}}$ can not be constant throughout the cluster as
assumed by Mroczkowski (2011). Indeed, such an assumption is
inconsistent with our other model assumptions as they lead to $f_{\rm {gas}}$ 
being a function of $r$. The profiles also show
a pronounced dependency on the cluster mass, reflecting the dependency
on the temperature as expected both from numerical simulations and
X-ray observations of galaxy clusters using XMM and \textit{Chandra} satellites
(Ettori et~al. 2004; Allen et~al. 2004; Sadat
et~al. 2005; Vikhlinin et~al. 2005, 2006; LaRoque et~al. 2006; McCarthy et~al. 2007; 
Afshordi et~al.2007).

Moreover, by numerically exploring the probability distributions of 
the cluster parameters given simulated interferometric SZ data in the context of 
Bayesian methods, and assuming our model with its corresponding assumptions, we 
investigate the capability of this model and analysis to return the simulated 
cluster input quantities. We find that simulated cluster physical 
parameters are well-constrained except $c_{200}$ which is relatively unconstrained. We can 
also recover the true values of the simulated clusters. In particular, the mean cluster total 
mass estimate $M_{\rm{tot}}(r_{200})$ and $r_{200}$ for the first simulated cluster are: 
$M_{\rm{tot}}(r_{200})=(5.1\pm 1.7)\times 10^{14}\,\rm{M_\odot}$ and $r_{200}=(1.5\pm0.2)\, 
\rm {Mpc}$ and the corresponding true values of the simulated cluster are: 
$M_{\rm{tot}}(r_{200})=5\times 10^{14}\,\rm{M_\odot}$ and $r_{200}=1.5\, \rm {Mpc}$. We 
determine the best-fit values of the parameters describing our model, i.e. 
$\rho_{\rm s}$, $R_{\rm s}$, $r_{\rm p}$ and $P_{\rm {ei}}$, and hence
calculate profiles of cluster total mass and gas pressure as determined using SZ data. We 
show that these profiles are consistent with our model.

We then repeat the analysis for a real cluster (A611) observed through its SZ effect with 
AMI. For this cluster, We find $M_{\rm {tot}}(r_{\rm 200})=(8.6\pm 
1.4)\times 10^{14}\,\rm{M_\odot}$ and $r_{200}=(1.7 \pm 0.1)\, \rm{Mpc}$. A611 has previously 
been studied in different wave-bands. For example, Schmidt \& Allen 
(2007) analysed \textit{Chandra} data of A611 and found  $M_{\rm {tot}}(r_{\rm 200})\approx 8 \times 10^
{14}\,\rm{M_\odot}$ and $r_{200}=1.7 \, \rm{Mpc}$. Donnarumma et~al. (2011) also studied 
\textit{Chandra} X-ray data of A611 with different assumptions on background and metallicity. Their estimates of the cluster total mass vary from $M_{\rm {tot}}(r_{\rm 200})=(9.32\pm 
1.39)\times 10^{14}\,\rm{M_\odot}$ for $r_{200}\approx1.8 \, \rm{Mpc}$ to $M_{\rm {tot}}(r_{\rm 200})=(11.11\pm 2.06)\times 10^{14}\,\rm{M_\odot}$ for $r_{200}\approx1.96 \, \rm{Mpc}$. They also carried out a strong lensing analysis of the 
cluster and found the mass estimates vary from  $M_{\rm {tot}}(r_{\rm 200})=(4.68\pm 
0.31)\times 10^{14}\,\rm{M_\odot}$ for $r_{200}\approx1.4 \, \rm{Mpc}$ to $M_{\rm {tot}}(r_{\rm 200})= 6.32_{-0.23}^{+0.51}
\times 10^{14}\,\rm{M_\odot}$ for $r_{200}\approx1.5 \, \rm{Mpc}$ when using different techniques. From weak lensing study of the 
cluster, Romano et~al. (2010) find that the cluster total mass within radius of $1.5 \rm {Mpc}
$ is $(8\pm 3)\times 10^{14}\,\rm{M_\odot}$ from the aperture mass technique and $(5\pm 1)
\times 10^{14}\,\rm{M_\odot}$ assuming parametric models. Our previous SZ analysis of A611 
using isothermal $\beta$ model (AMI Consortium: Shimwell et~al. 2011) resulted in $M_{\rm 
{tot}}(r_{\rm 200})=(5.7\pm 1.1)\times 10^{14}\,\rm{M_\odot}$ and $r_{200}=(1.6 \pm 0.1)\, \rm
{Mpc}$. Comparing the results of these studies with our analysis reveals that our results are 
in good agreement with the results of X-ray and weak lensing analyses of A611 but strong 
lensing and our previous SZ analyses of the cluster find a lower cluster mass which 
might be due to the extrapolating the strong lensing results in the outer spatial range 
as has been pointed out by Donnarumma et~al. (2011) and also the assumption of isothermality 
in our previous SZ study of the cluster. 

We conclude that our proposed simple model for spherical galaxy
clusters leads to realistic radial profiles for all the properties of
interest, and hence may prove useful in the analysis of
multi-wavelength cluster observations.  An obvious future avenue for
research, which we will explore in a follow-up paper to this letter,
is to iterate the solution we have obtained by inserting the derived
$\rho_{\rm {gas}}(r)$ in (\ref{eq:rhogas}) back into the expression
for the total density $\rho_{\rm {tot}}(r)=\rho_{\rm {DM}}(r) +
\rho_{\rm {gas}}(r)$, recalculating the form of the other variables
and repeating this process until convergence is established. In so
doing, one might hope to obtain an even more realistic cluster model,
but at the cost of losing a simple analytical formulation.
\section*{Acknowledgments}
The authors thank their colleagues in the AMI Consortium for numerous
illuminating discussions regarding the modelling of galaxy clusters.The data analyses were 
carried out on the COSMOS UK National Supercomputer at DAMTP, University of Cambridge and we 
are grateful to Andrey Kaliazin for his computing assistance. MO acknowledges an STFC 
studentship.
\setlength{\labelwidth}{0pt} 

\label{lastpage}
\end{document}